\pdfoutput=1
\documentclass[preprint, 11pt, authoryear]{elsarticle}

\usepackage{geometry}
 \usepackage{graphicx,booktabs,calc}
 
\usepackage{pifont}

\usepackage{lineno,hyperref}
\usepackage{hyperref}
\usepackage{multirow}
\usepackage{amssymb}
\usepackage{subcaption} 
\usepackage{xcolor}
\usepackage[colorinlistoftodos,prependcaption]{todonotes}
\usepackage[flushleft]{threeparttable}
\usepackage{enumitem}
\usepackage{makecell}
\usepackage{array}
\usepackage{adjustbox}
\usepackage{rotating}
\usepackage{amsmath,mathtools}
\usepackage{pdflscape}
\usepackage[font=small,labelfont=normalsize]{caption}

\journal{Simulation Modelling Practice and Theory}

\biboptions{authoryear}

\begin{document}

\begin{frontmatter}

\title{Network Impacts of Automated Mobility-on-Demand: A Macroscopic Fundamental Diagram Perspective}

\author[smart]{Simon Oh}
\author[smart]{Antonis F. Lentzakis}
\author[smart]{Ravi Seshadri}
\author[mitcee]{Moshe Ben-Akiva}
\address[smart]{Future Urban Mobility, Singapore-MIT Alliance for Research and Technology (SMART)
1 CREATE Way, \#09-02 CREATE Tower, Singapore 138602}
\address[mitcee]{Department of Civil and Environmental Engineering, Massachusetts Institute of Technology, Cambridge, MA 02139, United States}



%

\begin{abstract}
Technological advancements have brought increasing attention to Automated Mobility on Demand (AMOD) as a promising solution that may improve future urban mobility. During the last decade, extensive research has been conducted on the design and evaluation of AMOD systems using simulation models. This paper adds to this growing body of literature by investigating the network impacts of AMOD through high-fidelity activity- and agent-based traffic simulation, including detailed models of AMOD fleet operations. Through scenario simulations of the entire island of Singapore, we explore network traffic dynamics by employing the concept of the Macroscopic Fundamental Diagram (MFD). Taking into account the spatial variability of density, we are able to capture the hysteresis loops, which inevitably form in a network of this size. Model estimation results at both the vehicle and passenger flow level are documented. Environmental impacts including energy and emissions are also discussed. Findings from the case study of Singapore suggest that the introduction of AMOD may bring about significant impacts on network performance in terms of increased VKT, additional travel delay and energy consumption, while reducing vehicle emissions, with respect to the baseline. Despite the increase in network congestion, production of passenger flows remains relatively unchanged.
\end{abstract}


\end{frontmatter}


\section{Introduction}
Recent technological advancements are changing the way we view urban mobility systems and are set to bring about a host of opportunities to improve mobility, accessibility, and livability. This is evident from the advent of transportation networking companies (TNC) and ride-sourcing services, hereafter termed  Mobility-on-Demand (MOD). 
TNCs are rapidly embracing new business models of shared mobility, on-demand ride-hailing and seamless multimodality, by employing a multi-sided business platform which attracts both drivers and customers (passengers). App-based MOD services have become an entrenched mobility option penetrating 7-8\% of the market, generating 44 billion USD of worldwide revenue in 2017 \citep{oecdreport2018}, and are projected to reach a market penetration rate of 13\% with double the revenue within five years \citep{statistareport2018}. The main factors to which the large adoption rates can be attributed are respondents' satisfaction with low waiting and travel times, ease-of-use, and the convenience of smartphone-based services \citep{rayle2016just}.

The potential of integrating autonomous vehicle (AV) technology and ride-sourcing platforms, as part of AV-based on-demand shared-ride services, hereafter termed Automated Mobility-on-Demand (AMOD), has been well recognized by major technology companies. Significant progress has been made in AV technology itself by the traditional automotive industry as well as the emerging AV software platform companies, including Nvidia Drive AGX, Aptiv (formerly Delphi Connection Systems), Waymo (formerly Google Self Driving Car project). Technology companies have been running trials on AV-based mobility services, e.g., Waymo has accumulated more than 10 million miles of on-road testing from 2009 to 2018. Some major players are contributing to the realization of AMOD services by entering into partnerships with traditional car-makers and TNCs, e.g. the Early Ride Program by Waymo with self-driving Chrysler cars in Phoenix, the first commercial service by Aptiv which takes advantage of the ride-hailing network of Lyft with an autonomous fleet of BMW cars in Las Vegas.

Recent market research \citep{alliedmarketresearch2018} projects the growth of the global autonomous mobility market to increase from 5 billion USD (in 2019) to 556 billion USD (in 2026) with foreseeable benefits including improved safety (given the fact that 94\% of accidents are caused by human factors), higher transportation network throughput, improved efficiency (with centralized fleet operation), more affordable services (due to competitive cost structures), as well as other long-term benefits on urbanization. However, these benefits are as of yet far from guaranteed, because of economic and social barriers (\cite{fagnant2015preparing}), large uncertainty on the cost and pricing of AMOD (\cite{bosch2018cost}), and potential adverse effects of AMOD on existing transportation systems, such as induced demand, cannibalization of transit, congestion, increased Vehicle-Kilometers-Traveled (VKT), and empty trips involving dead heading (\cite{simoni2019congestion}; \cite{horl2019fleet}; \cite{zhang2018impact}), as has already been observed with MOD services (\cite{WashingtonPost}). For this reason, a recent white paper \citep{trbwhite2018} also points out the importance of studying the design of AMOD systems (involving fleet management and operation, supply of infrastructure for charging and parking) and their impacts on transportation (including system capacity, VKT, transit, travel behavior and land use patterns). Regarding future challenges, the standing committee on traffic flow theory and characteristics (TFTC) suggests specific directions over four primary areas:  simulation, connected and automated vehicle technologies, network-wide modeling, and multimodality (\cite{ahn2019traffic}).

In this respect, this paper studies the potential network impacts of AMOD using an agent- and activity-based traffic simulation platform. Demand is modeled using an activity-based model system (ABM), that draws on stated preferences data from a smartphone-based survey in Singapore. Supply is modeled using an on-demand mobility service controller (that replicates the operations of MOD/AMOD fleets involving assignment and rebalancing of service vehicles) integrated within a mesoscopic multimodal network simulator. Interactions between demand and supply are explicitly modeled. Through scenario simulations of the entire network of Singapore, we contribute to the literature on AMOD, by employing network-wide Macroscopic Fundamental Diagrams (hereafter termed as MFDs) to explore congestion patterns over the entire network. In order to examine the impact of introducing AMOD services on existing multimodal networks, we take inspiration from past literature on generalization (e.g. \cite{ramezani2015dynamics}) and extension of the MFD concept (e.g. \cite{geroliminis2014three}).

\section{Past Research}\label{sec:PreviousResearch}
\subsection{AMOD System Design and Evaluation}\label{sec:PreviousResearch_AMOD}
Extensive research, employing simulation-based optimization methods, has endeavoured to analyze the impact of AMOD services on transportation networks. Initial studies examined the potential of AMOD services using queuing theory and network models. \cite{spieser2014toward} estimated the AMOD fleet size required to serve all existing private vehicle trips in Singapore and concluded that fewer vehicles are required to serve existing demand with reasonable waiting times. Along similar lines, \cite{burns2015transforming} analyzed travel patterns, cost estimates, and vehicle requirements for different network configurations corresponding to mid-sized, low-density, and densely-populated urban areas. 

Researchers have also addressed the deployment and operations of on-demand services and proposed novel vehicle assignment and rebalancing strategies to efficiently deal with spatio-temporal variations in demand. Linear and integer programming approaches were utilized for the minimization of vehicle rebalancing while maintaining vehicle availability over the network (\cite{pavone2011load}, \cite{zhang2016control}). Similarly, \cite{zachariah2014uncongested} solved a rebalancing assignment problem of AV taxis in New Jersey by minimizing the number of empty vehicles on the network. Researchers have also proposed solutions to the fleet sizing problem using the concept of shareability networks and --using the New York taxicab dataset-- have shown a significant reduction in the cumulative trip length (\cite{santi2014quantifying}) and required fleet size to accommodate existing demand (\cite{alonso2017}; \cite{vazifeh2018}). \cite{hyland2018} employed an agent-based simulation, which uses a mathematical programming solver to compare a variety of heuristic and optimization-based assignments in grid networks. Presenting a case study with Chicago taxi demand data, they suggest that ‘sophisticated’ assignment algorithms are able to serve more incoming requests with limited fleet size and result in fewer empty vehicles within the fleet.

Regarding the effects of AMOD services, \cite{martinez2017}, using agent-based simulations, reported the potential reduction of vehicle population, travel volume, and parking spaces and increased fleet mileage in Lisbon, Portugal. Similar findings have also been reported in \cite{fagnant2014travel}, who examined AMOD service impacts with a portion of existing trips (taken from NHTS, 2009) in a synthetic city similar to Austin, Texas. Their results showed that shared AVs (hereafter termed as SAVs) can fulfill the vehicle needs of nearly 12 privately owned cars, serve 31 to 41 requests per  day, and reduce the required parking spaces by 11 per service vehicle. However, these  studies
fail to capture network congestion effects, as well as the interactions between demand and supply. 

Recent studies have addressed the aforementioned shortcomings using agent-based traffic simulations. \cite{boesch2016autonomous} determined the fleet sizes required to satisfy different levels of demand in the greater Zurich area, Switzerland, using the multi-agent transport simulation software MATSim (\cite{horni2016multi}) and reported that a significant reduction in the vehicle population can be achieved when introducing an AMOD service (that can fulfill requests within a waiting time of 10 minutes, similar to previous literature). \cite{bischoff2016simulation} obtained similar results on the replacement of private trips, for the city of Berlin, by solving the dynamic vehicle routing problem (DVRP) with MATSim. \cite{maciejewski2016congestion} investigated congestion effects of AV taxis with travel time and delay ratios for scaled-down scenarios over different settings (of replacement rates, fleet sizes, and road capacity levels) and suggested that large fleets may aggravate congestion because of unoccupied trips, assuming there is no road capacity improvement by automation. Further, simulation scenarios of Zurich from \cite{horl2019fleet} tested different AMOD fleet operational policies using the daily travel patterns extracted from a synthetic Swiss population (which generated around 360k trips for AMOD). The study reported that --using a feedforward fluidic rebalancing algorithm-- a fleet size of 7,000 vehicles was able to serve 90\% of requests within 5 minutes, and further examined the cost implications of AMOD services based on \cite{bosch2018cost}.
From a recent case study (\cite{segui2019simulating}) in Greenwich, London, UK, the authors integrated a fleet simulation software called IMSim to MATSim in order to evaluate different configurations of vehicle specifications, fleet sizes, parking and charging infrastructure and service criteria from traveler, operator, and city’s perspectives. The authors indicated the negative effects of AMOD, whereby AMOD fleet vehicles come to have additional travel distances, which may result in added congestion, thus emphasizing the need for future research to conduct more detailed investigations. In order to explicitly consider demand-supply interactions, \cite{azevedo2016microsimulation} analyzed the sensitivity of AMOD supply (i.e. fleet sizes, parking configurations) on travel behavior (i.e. mode shares, routes, and destination choices), and more recently, \cite{basu2018} investigated the potential of AMOD services to substitute mass transit, using an agent- and activity-based simulation platform. 


Despite the growing body of literature on AMOD systems, several limitations remain: 
\begin{enumerate}[label=(\roman*)]
    \item Simplified abstraction of the urban network including grid type networks (\cite{fagnant2014travel}), Euclidean planes (\cite{spieser2014toward}), quasi-dynamic grid-based networks (\cite{zhang2015queueing}; \cite{martinez2017}; \cite{fagnant2018dynamic}), synthetic grids (\cite{hyland2018}), prototypical cities (\cite{basu2018})
    \item Coarse-grained simulation models where approximations are made by teleporting trips between locations with static travel times and without using models of network congestion (\cite{spieser2014toward}; \cite{alonso2017}; \cite{fagnant2018dynamic}; \cite{farhan2018impact}; \cite{chen2016operations}; \cite{burns2015transforming}; \cite{zhang2016control}; \cite{boesch2016autonomous})
    \item Substituting a proportion of existing private trips with AMOD and limited modeling of behavioral preferences towards AMOD  (\cite{burns2015transforming}; \cite{boesch2016autonomous}; \cite{zhang2016control}; \cite{maciejewski2016congestion}; \cite{bischoff2016simulation}; \cite{horl2019fleet}). 
\end{enumerate}

To overcome these limitations, recent studies have started to integrate on-demand service simulators with a traffic simulator (i.e. \cite{segui2019simulating}; \cite{oh2020amod_tra}) to capture future impacts of AMOD on demand and supply. However, an analysis of network traffic dynamics has, to the best of our knowledge, not been conducted on large-scale urban networks, and consequently, the understanding of the network effects of AMOD still warrants investigation.

\subsection{Network-wide Traffic Modeling}\label{sec:PreviousResearch_MFD}
A recent trend for capturing congestion patterns of urban areas is modeling and analyzing network traffic dynamics at the urban-scale, utilizing the MFD concept. In the past decade, the spatial scale of traffic modeling has been extended to the network level, whereby aggregated traffic dynamics are described collectively over the urban area. Initial studies on macroscopic relationships dating back to the 1960s, determined the optimum density necessary for sustaining maximum flow rate in a given area (\cite{smeed1967road}; \cite{godfrey1969mechanism}). Following that, \cite{herman1979two} proposed a two-fluid model that models the relationship between average vehicular speed and density, later verified by simulation (\cite{mahmassani1987performance}). 
The concept of the MFD was formalized by assuming a homogeneous congestion distribution over an urban area (\cite{daganzo2007urban}) and empirically evidenced by the well-defined macroscopic relationship between network production (i.e. average flow, trip completion rate) and accumulation (average density, total vehicles on the network), in a study of Yokohama, Japan (\cite{geroliminis2008existence}). The existence of MFDs have since been verified and reproduced for other cities all over the world: Toulouse, France (\cite{buisson2009exploring}), Zurich, Switzerland (\cite{ambuhl2017empirical,loder2017empirics}), Rome, Italy (\cite{bazzani2011towards}), Sendai, Japan (\cite{wada2015empirical}), Shenzhen, China (\cite{ji2014empirical}), Brisbane, Australia (\cite{tsubota2014macroscopic}), Minnesota, USA (\cite{geroliminis2011hysteresis}), Amsterdam, Netherlands (\cite{knoop2013empirics}), Lyon, France (\cite{mariotte2018dynamic}).

The MFD concept has been employed in the implementation of large-scale traffic control measures by reducing vehicle accumulation to its critical level so as to mitigate overall congestion. It includes perimeter control, whereby metering of the number of vehicles into a specific ``protected" region takes place (\cite{daganzo2007urban}; \cite{haddad2012stability}; \cite{haddad2013cooperative}; \cite{keyvan2012exploiting}; \cite{ramezani2015dynamics}; \cite{geroliminis2012optimal}; \cite{kouvelas2017enhancing}; \cite{kim2018agent}), pricing affecting travel behavior on mode and destination choice (\cite{geroliminis2009cordon}; \cite{gonzales2012morning}; \cite{zheng2012dynamic}; \cite{simoni2015marginal}; \cite{zheng2016modeling}), route guidance (\cite{yildirimoglu2015equilibrium, lentzakis2018region}), space allocation (\cite{zheng2013distribution}), and parking (\cite{leclercq2017dynamic}).

To estimate the MFD, researchers have utilized both analytical and experimental approaches. \cite{daganzo2008analytical} analytically presented the `cuts method' based on variational theory by determining the different upper bounds on the MFD plane. Later, \cite{leclercq2013estimating} utilized this approach in estimating the MFD in simple networks with different routes, and \cite{laval2015stochastic} proposed a stochastic approximation method to estimate the MFD of an urban corridor based on variational theory. Studies employing experimental approaches estimated the flow and density with sensor data observed based on Eulerian (\cite{shoufeng2013deriving}) and Lagrangian (\cite{nagle2013method}) approaches. Readers can refer to \cite{leclercq2014macroscopic} for more details.

The shape of MFDs can be affected by several factors including network supply (e.g. geometric features, signal timings, road capacity, heterogeneity of congestion) and demand (e.g. route choice, detouring, OD flows). \cite{buisson2009exploring} attributed the loop-like hysteresis shape of the MFD to the local heterogeneity of sensor distribution over the network, network composition involving road types and spatial distribution of demand and local congestion, and were the first to relax the homogeneity conditions of the MFD described in earlier studies (\cite{geroliminis2008existence,geroliminis2007macroscopic}). This hysteresis phenomenon has been repeatedly observed or reproduced from further studies on both empirical data and simulation data (\cite{mazloumian2010spatial}; \cite{gayah2011clockwise}, \cite{daganzo2011macrore}, \cite{geroliminis2011hysteresis}, \cite{mahmassani2013urban}, \cite{muhlich2014examination}, \cite{saeedmanesh2015empirical}) showing different average flow rates during the onset and dissipation of congestion. In addition, the degree of spatial variation of network occupancy has been used to explain the size of hysteresis (\cite{saberi2012exploring,saberi2014estimating}). To incorporate the spatial variation into the MFD modeling framework, \cite{knoop2015traffic} generalized the MFD (GMFD), describing the relation between average flow with average density and density heterogeneity. The authors explained the occurrence of hysteresis as a result of nucleation effects and demonstrated the performance loss due to spatial heterogeneity. \cite{knoop2013empirics} predicted network production by formulating the GMFD with both non-parametrized and parameterized forms. \cite{ramezani2015dynamics} also integrated the dynamics of heterogeneity into the aggregated model for subregion-based MFDs and their perimeter control.

The effect of route choice behavior on the scatter of MFD has been explored by many studies (\cite{yildirimoglu2015equilibrium,leclercq2013estimating,gayah2011clockwise,gayah2014impacts}). \cite{leclercq2013estimating} posited that the scatter of MFD is affected by route choices and (uneven/inconsistent) distribution of congestion. \cite{gayah2011clockwise} showed in simulations that hysteresis loops can be reduced in size through adaptive route choice with respect to congestion. Also, demand patterns (derived from route choice) have been identified as a factor leading to bifurcation at the high density part of MFD (\cite{leclercq2015macroscopic,shim2019empirical}) and network instability (\cite{daganzo2011macrore,mahmassani2013urban}).

Recent studies have extended the MFD into three dimensions to explain the passenger and vehicle flow in multimodal networks. One notable study by \cite{geroliminis2014three} suggests a three-dimensional MFD capturing the performance of bi-modal networks by relating the accumulation of cars and buses with the vehicle and passenger flow, which they call 3D-vMFD, 3D-pMFD respectively.  \cite{ampountolas2017macroscopic} proposed a solution to the perimeter control problem by controlling the vehicle composition of bi-modal traffic. \cite{loder2017empirics} was able to derive 3D-MFDs using data from loop detectors and public transit in the city of Zurich. The authors estimated the 3D model using a linear relationship between vehicle density and speed for each mode and measured the effect of vehicle accumulation on the speed of cars and buses. These studies suggested negative marginal effects for additional vehicles (higher for bus than car) on network speed. \cite{paipuri2020bi} simulated three different MFD-based models (accumulation-, trip- and delay accumulation-based approach) over different traffic states considering the 3D-MFD concept for a grid network with dedicated bus lanes. The authors highlighted the importance of segregated 3D-MFDs to accurately resolve traffic dynamics. 

In summary, extensive research has been conducted in regard to both AMOD system design and the modeling of network-wide traffic. However, despite the extensive literature, the network impact of AMOD services, with respect to congestion, still warrants further investigation, particularly in large-scale urban networks. This paper attempts to fill the gap between these two areas by explicitly investigating network-wide congestion effects from the MFD perspective through a high-fidelity agent-based traffic simulation platform. Following this section, Section \ref{sec:SecMethodology} presents the agent-based simulation framework and the formulation of the MFD for the simulation scenarios described in Section \ref{sec:SecScenario}. Then, in Section \ref{sec:SimResult} we analyze and estimate the network-wide MFD (Section \ref{sec:Result_MFD}), followed by, Section \ref{sec:Result_Effects}, which discusses the impacts of congestion from the  standpoint of traveler, operator, and planner. Finally, Section \ref{sec:Conclusion} presents conclusions, as well as future research directions.

\section{Methodology}\label{sec:SecMethodology}
\subsection{Simulation Framework}\label{sec:SecSimMethod}
\begin{figure}[ht]
\centering
 \includegraphics[scale=0.55]{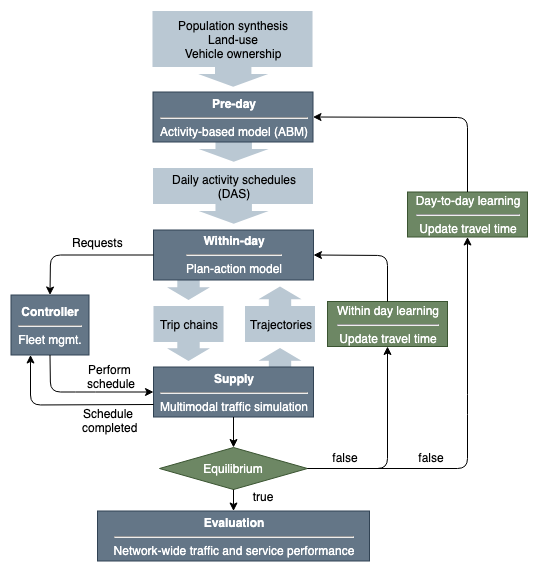}
\caption{Simulation Framework}\label{fig:Fig1}
\end{figure}

We utilize the high-fidelity activity- and agent-based simulation platform (\textit{SimMobility} (\cite{adnan2016simmobility}) to model daily network-wide trips, for all agents in an urban area. \textit{SimMobility} is comprised of three primary components operating at different temporal scales, the Short-term, Mid-term and Long-term. 
In this study, we will primarily make use of
\textit{SimMobility Mid-term} (\cite{lu2015simmobility}), which models daily activity and travel demand and simulates multimodal network performance at a mesoscopic level.
The Mid-term is composed of three modules, the \textit{Pre-day}, \textit{Within-day}, and \textit{Supply}, as shown in Figure \ref{fig:Fig1}. 

The \textit{Pre-day} module is a system of hierarchical discrete choice models (logit and nested-logit) and simulates the daily activity patterns of individuals through an activity-based model system (\textit{ABM}) \citep{ben1996travel}. The pre-day model system consists of three levels: 

\begin{itemize}
\item	The day pattern level generates a list of tours and availability of intermediate stops for each activity type (work, education, shopping, and others).
\item	The tour level describes the details for each tour including destination, travel mode, time of day (arrival time and departure time) and occurrence of work-based sub-tours. 
\item	The intermediate stop level generates the intermediate stops for each tour and predicts the details of the secondary activities (including destination, mode, etc).
\end{itemize}

The \textit{Pre-day} model system provides the daily activity schedule (\textit{DAS}) – a detailed description of individual activity and mobility patterns, including arrival/departure time, destination (at zonal level), and travel mode for each trip/tour. Interested readers can find more details of the \textit{Pre-day} model in \cite{siyu2015activity}.

At the \textit{Within-day} level, the pre-day activity schedule is transformed into actions by performing departure time choice, route choice and within-day re-scheduling of individual trips \citep{ben2010planning}. Following this, the \textit{Supply} module simulates network dynamics using macroscopic traffic flow relationships (speed-density models) combined with deterministic queuing models, as well as public transit operations through bus and rail controllers that dispatch vehicles (frequency/headway-based operation), monitor the vehicle occupancy, and determine the dwell time at stops/stations. The \textit{Supply} model also includes a \textit{Smart Mobility Service (SMS) controller} that replicates the operations of an on-demand ride-sharing mobility service (\cite{basu2018}). For trips that require on-demand services (MOD, AMOD), the agent (passenger) sends a ride request to the controller with pertinent details, including service type (single, shared), and origin/destination for Pick-Up/Drop-Off (PUDO). Subsequently, the controller accommodates the client’s request by assigning and dispatching the service vehicle from the available vehicle list in the fleet which satisfies constraints on:
\begin{enumerate}[label=(\roman*)]
    \item new passenger’s minimum waiting time
    \item existing passenger’s additional travel time due to detours
    \item the number of seats available in the service vehicle.
\end{enumerate}

When idle, vehicles are directed to i) cruise within a specific area (i.e. high demand zone) or ii) drive to a parking location (i.e. the nearest available) until the controller finds a new request to assign to the vehicle. 

In order to ensure equilibrium (or consistency) between demand and supply, after running the \textit{Supply} simulation for a given scenario, we iteratively adjust the travel time tables (comprising of link travel times and public transit waiting times). The objective of the within-day learning process is to achieve equilibrium with regard to route choice decisions. Specifically, we compute the travel time in iteration $i+1$ ($t_{i+1}$) as a weighted sum of the current travel time from the supply simulation ($t_{S}$) and travel time in iteration $i$ ($t_{i}$): $t_{i+1}=t_{i}*w+t_{S}*(1-w)$, where $w$ is a parameter. This process is repeated until the travel times in successive iterations ($t_{i+1}$, $t_{i}$) converge. Similarly, the day-to-day learning process enables the \textit{Pre-day} model system to adjust the individual activity schedules with updated travel times (including zone-to-zone travel-times, waiting times for public transit and waiting times for MOD and AMOD services). This process allows for the re-evaluation of accessibility, using agents’ actual travel-times, experienced during the \textit{Supply} simulation and arrive at a `day-to-day' equilibrium.

\subsection{Network Performance}\label{sec:SecMethod_model}
As noted previously, the multimodal \textit{Supply} simulation provides detailed information of individual agent and service vehicle trajectories. Travel trajectories contain information about the departure/arrival time at origin/destination, travel distance, and travel mode of each individual agent. Service vehicle trajectories contain information regarding schedule items performed by each service vehicle and their status in each time interval. These trajectories allow us to estimate network-wide traffic measures. 

Network performance of each scenario is evaluated using suitable macroscopic variables, as detailed subsequently. Density is measured at the segment level ($k_n$ for segment $n$) across the network and vehicle accumulation ($\mathcal{A}_V$, unit: veh; note that the subscript $V$ denotes vehicles and $P$ denotes passengers) is given by : 

\begin{equation}\label{eq:eq_Av}
\mathcal{A}_V = \frac{\sum_{n}^{N_s}k_n \cdot l_n}{\sum_{n}^{N_s}l_n} \cdot L_N
\end{equation}

Where, $l_n$ is the length of segment $n$; $L_N$ is the total network length. $N_s$ represents the number of segments equipped with sensors and is a subset of the total number of segments $N$. While $N_s$ would be useful from a practical implementation perspective, in this paper, data from all links are made available to us ($N_s = N$). The resulting accumulation may also be expressed as the sum of accumulations of each mode (at the vehicle level):

\begin{equation}
\mathcal{A}_V = \sum_{v \in \mathcal{V}}\mathcal{A}_v
\end{equation}

Where, $\mathcal{V}$ denotes the set of road-based modes. Also note that the spatial density variability ($\gamma$, unit: veh/km) is measured using the standard deviation of segment density ($k_n$) as in Eq. \ref{eq:eq_spread_density}. Vehicle production ($\mathcal{P}_V$, unit: veh-km/hr) represents the total travel distance ($VKT$) driven by vehicles per unit time which can be quantified using the flow at each segment $q_n$:

\begin{equation}\label{eq:eq_Pv}
\mathcal{P}_V = \frac{\sum_{n}^{N_s}q_n \cdot l_n}{\sum_{n}^{N_s}l_n} \cdot L_N
\end{equation}

As noted previously, the travel trajectories capture detailed information of the mobility pattern of each individual vehicle/passenger including departure time, origin/destination, activity details (type and duration), travel (waiting) times, and average trip distances ($TD_V$, $TD_P$). 
Information is also available for respective trip completion rates ($TC_V$ and $TC_P$, unit: trips/hr) that provide the number of completed trips per unit time. The production of passenger flow ($\mathcal{P}_P$) is thus estimated using the trip completion rate ($TC_P$) and average trip distance ($TD_P$) at the passenger level as,  

\begin{equation}\label{eq:eq_Pp}
\mathcal{P}_P = \sum_{p \in \mathcal{P}}TC_p \cdot TD_p
\end{equation}

Where, $\mathcal{P}$ denotes the set of all passenger modes. Equation \ref{eq:eq_Pp} allows us to accurately measure production of passenger flow without the need to use average passenger occupancy as is typically done (\cite{geroliminis2014three,ampountolas2017macroscopic,loder2017empirics}).  
The number of travelers in the simulation (captured at each time interval over the entire network) represents the passenger accumulation ($\mathcal{A}_P$). Modes at the vehicle ($V$) and passenger level ($P$) are summarized in Table \ref{Table_TripSupportModes} in Section \ref{sec:Result_MFD}.

With this background, the MFD expresses the network production ($\mathcal{P}$) as a function of accumulation ($\mathcal{A}$) and congestion heterogeneity ($\gamma$) as in the literature (i.e. \cite{knoop2013empirics}; \cite{ramezani2015dynamics}),

\begin{equation}\label{eq:Eq_mfd}
\mathcal{P} = f(\mathcal{A}, \gamma)
\end{equation}

The heterogeneity term $\gamma$ typically refers to the spatial spread of density:

\begin{equation}\label{eq:eq_spread_density}
\gamma = \sqrt{\frac{\sum_{n}^{N}(k_n-\overline{k})^2}{N-1}}  
\end{equation}

MFD-based models have been extended to address congestion heterogeneity, as well as multimodality in various networks as described in Section \ref{sec:PreviousResearch_MFD}. In this paper, we adapt the exponential form found to be applicable to multimodal traffic (\cite{geroliminis2014three}) as well as heterogeneous urban networks (\cite{ramezani2015dynamics}). This approach formulates the \textit{vMFD} and \textit{pMFD}, corresponding to vehicles and passengers, as:

\begin{equation}\label{eq:eq_vmfd_hete}
\mathcal{P}_V(A_V, \gamma) = a \cdot \mathcal{A}_{V} \cdot e^{b\mathcal{A}_{V}^3+c\mathcal{A}_{V}^2+d\mathcal{A}_V+r\gamma}
\end{equation}

\begin{equation}\label{eq:eq_pmfd_hete}
\mathcal{P}_P(\mathcal{A}_V, \gamma, \mathcal{A}_P) = a \cdot \mathcal{A}_{V} \cdot e^{b\mathcal{A}_{V}^3+c\mathcal{A}_{V}^2+d\mathcal{A}_V+r\gamma+\rho\mathcal{A}_P}
\end{equation}

where $a, b, c, d, r, \rho$ are model parameters. 




\section{Scenarios}\label{sec:SecScenario}
\begin{figure}[ht]
\centering
 \includegraphics[scale=0.55]{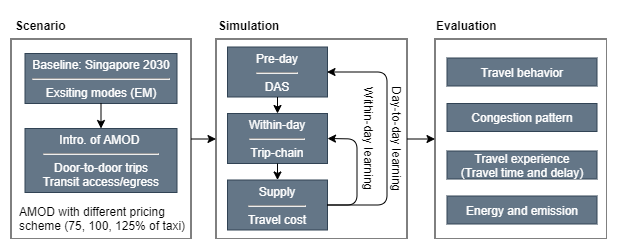}
\caption{Evaluation Framework}\label{fig:Fig_SceFrame}
\end{figure}

\begin{figure}[ht]
\centering
 \includegraphics[scale=0.5]{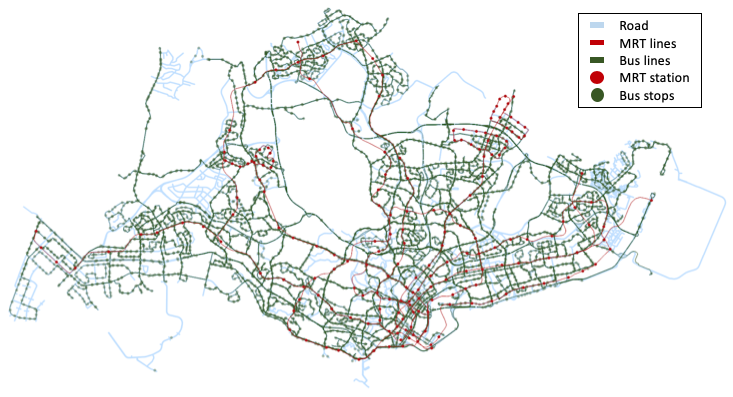}
\caption{Network Topology in Singapore}\label{fig:Fig_Network}
\end{figure}

The simulation scenarios in this study utilize a model of Singapore for the year 2030. The synthetic population of individuals and households (that are the inputs to the SimMobility Mid-term simulator shown in Figure \ref{fig:Fig_SceFrame}) were generated by a Bayesian network approach (\cite{sun2015bayesian}; details of the synthetic population can be found in \cite{oh2020amod_tra}). The network (Figure \ref{fig:Fig_Network}) consists of 1,169 traffic analysis zones, 6,375 nodes, 15,128 links, and 30,864 segments. The total network length ($L_N$) is approximately 3,175km, and includes 730 bus lines serving 4,813 bus stops, and 26 MRT (rail) lines serving 186 stations. 

Travel and activity demand is estimated by the \textit{Pre-day} ABM system using the synthetic population for year 2030 (for more details on estimation and calibration of the ABM system refer to \cite{oh2020amod_tra}) and also draws on data from a smartphone-based state preferences (SP) survey on AMOD (\cite{seshadri2019}). Three scenarios are considered with regard to the price or fare of the AMOD services:

\begin{itemize}
    \item AMOD single-ride price: 75\%, 100\% and 125\% of taxis
    \item AMOD shared-ride price: 75\% of single-ride 
\end{itemize}

Note that the taxi fare ($f_{taxi}$, unit: SGD) is determined as:
\begin{equation}\label{eq:eq_taxi_fare}
    f_{taxi} = f_{base} + f_{km} + f_{min}
\end{equation}
In which, $f_{base} = 3.2$, $f_{km} = 0.55 (<10km), 0.63 (>10km)$ per unit km, and $f_{min} = 0.29$ per min.

Thus, we simulate four scenarios of interest that differ in modal availability and AMOD pricing: Baseline, and three AMOD scenarios with different pricing schemes (75\%, 100\% and 125\% of taxis). In the baseline, travel modes available to agents are the existing modes (EM), which include private car, car-pooling (with 2 or 3 people per household), private bus, walking, taxi, MOD (Uber-like ride-sourcing services), public transit (bus, rail) with access/egress by walk. In the AMOD scenarios, in addition to the existing modes, the AMOD service is made available to travelers. AMOD services include door-to-door services with single/shared rides and first/last-mile connectivity to public transit (e.g. rail station). 

\begin{figure}[ht]
\begin{minipage}{0.5\textwidth}
  \centering
    \includegraphics[scale=0.45]{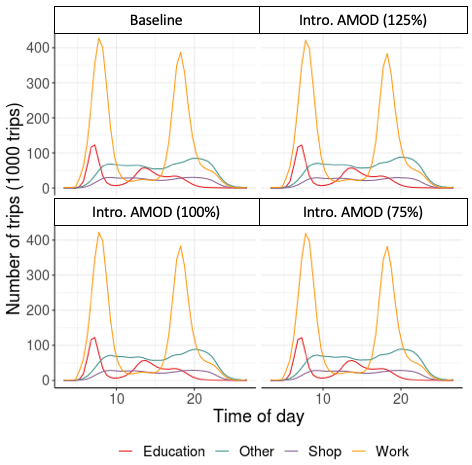}
	\subcaption{By Activity Type}\label{fig:Fig_Demand_Type}
\end{minipage}
~\hfill
\begin{minipage}{0.5\textwidth}
  \centering
    \includegraphics[scale=0.45]{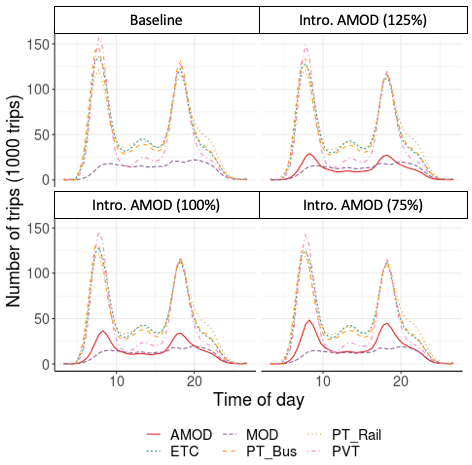}
	\subcaption{By Mode}\label{fig:Fig_Demand_Mode}
\end{minipage}
\caption{Travel Demand Pattern over Time-of-day}\label{fig:Fig_Demand}
\end{figure}

Figure \ref{fig:Fig_Demand} shows the distribution of demand for the different pricing scenarios by mode and activity types, each of which shows a different temporal pattern (Figure \ref{fig:Fig_Demand_Type}). The \textit{Work} trips comprise the largest portion of trips particularly during the peak periods. \textit{Education} trips show similar patterns with \textit{Work} in the morning, however, as expected, many trips occur before the PM peak period (around 2--3PM). Trips for \textit{Shopping} and \textit{Other} activities (such as leisure, recreation) are observed throughout the day. A large number of additional trips for \textit{Other} activities occur during and after PM peak. 

\begin{table*}[ht]
\centering
\begin{threeparttable}
\caption{Mode Share}\label{Table_ModeShare}
\begin{tabular}{l l c c c c}
\hline
\multirow{2}{*}{\textbf{Modes}} & & \multirow{2}{*}{\textbf{Baseline}} & \multicolumn{3}{c}{\textbf{Intro. of AMOD}} \\
\cmidrule(r){4-6}
& & & 75\% & 100\% & 125\% \\
\Xhline{1.5pt}
\multirow{2}{*}{PVT} & Car/Carpool & \makecell{18.75\%} & \makecell{17.33\%} & \makecell{17.7\%} & \makecell{17.93\%} \\
\cline{2-6}
& Taxi & \makecell{2.16\%} & \makecell{1.6\%} & \makecell{1.69\%} & \makecell{1.75\%} \\
\hline
\multirow{4}{*}{PT} & Bus & \makecell{24.33\%}	& \makecell{21.49\%} & \makecell{22.14\%} & \makecell{22.57\%} \\
\cline{2-6}
& Rail(Walk)\tnote{a} & \makecell{23.81 \%} & \makecell{20.54 \%}	& \makecell{21.21\%}	& \makecell{21.67\%} \\
\cline{2-6}
& Rail(MOD)\tnote{a} & \makecell{0.36\%} & \makecell{0.3\%} & \makecell{0.32 \%} & \makecell{0.32\%} \\
\cline{2-6}
& Rail(AMOD)\tnote{a} & 0 & \makecell{2.31\%} & \makecell{1.88\%} & \makecell{1.55\%} \\
\hline
MOD & Single/Shared & \makecell{6.41\%} & \makecell{5.38\%} & \makecell{5.51\%} & \makecell{5.64\%} \\
\hline
AMOD & Single/Shared & 0 & \makecell{8.87\%} & \makecell{7.01\%} & \makecell{5.77\%} \\
\hline
Other & & \makecell{24.16\%} & \makecell{22.18\%} & \makecell{22.54\%} & \makecell{22.79\%} \\
\hline
\end{tabular}
\begin{tablenotes}
    \small
    \item[a] Access/egress to/from rail station by Walk, MOD, and AMOD respectively.
\end{tablenotes}
\end{threeparttable}
\end{table*}

Table \ref{Table_ModeShare} lists the mode shares for each scenario (temporal distribution in Figure \ref{fig:Fig_Demand_Mode}). The total number of passenger trips for 24 hours is 8,991,057 trips (baseline), 8,995,544 trips (75\% pricing), 8,992,168 trips (100\% pricing), 8,994,926 trips (125\% pricing). These passenger trips (around 9 million) for all scenarios are simulated along with background traffic of freight vehicles (665,929 trips) estimated by the SimMobility Freight model (\cite{sakai2019modeling}). As expected, the introduction of AMOD leads to a reduction in the share of existing modes. Particularly, the share of public transit (PT), including Bus and Rail, reduces by 2.39--3.86\%, while reductions in the number of private vehicle trips (PVT) are smaller in magnitude (1--2\%). Thus, a large portion of AMOD demand includes shifts from PT (more than 55\%), while the shift rates from other modes are relatively low (around 4\%, 14\%, 5\% of AMOD demand are from private car, taxi, and MOD trips, respectively). Overall, the shares of AMOD range from 5.77--8.87\% across the three pricing scenarios. 

Public transit vehicles (buses and trains) operate in accordance with fixed schedules as described in Section \ref{sec:SecSimMethod}. Regarding the on-demand services, the fleet sizes for the three AMOD pricing scenarios (75\%, 100\% and 125\% respectively) are fixed at 43,000, 33,000, and 27,000 vehicles comprising 4- and 6-seaters (see \cite{oh2020amod_tra} for more details). Note that this fleet size is derived by finding an \textit{optimal} size, which yields sufficient fleet utilization (minimizing the number of idle vehicles during peak period), reasonable passenger waiting times (less than 6 min) and service satisfaction rates (serving all incoming requests). The required MOD fleet size ranges from 20,000--22,000 for each scenario. 


Table \ref{Table_Settings} summarizes the simulation configurations and scenario factors described in this section. Each scenario was simulated via several iterations of the \textit{within-day} and \textit{day-to-day learning} process to ensure the consistency between demand and supply.


\begin{table*}[ht]
\centering
\begin{threeparttable}
\caption{Experimental Settings}\label{Table_Settings}
\begin{tabular}{c l | p{1.5cm} | p{1.5cm} | p{1.5cm} | p{0.7cm} | p{0.7cm}}
\hline
    \multirow{3}{*}{\textbf{Factor}} & & \multicolumn{5}{c}{\textbf{Scenarios}} \\
    \cline{3-7}
    & & \multirow{2}{*}{\textbf{Baseline}} & \multicolumn{4}{c}{\textbf{Intro. AMOD}}\\
    \cline{4-7}
    & & & 75\% & 100\% & \multicolumn{2}{c}{125\%} \\    
\Xhline{1.5pt}
    \multirow{3}{*}{\makecell{Simulation \\ config.}} & Simulation model & \multicolumn{5}{c}{SimMobility Mid-term} \\
    \cline{2-7}
    & Simulation period & \multicolumn{5}{c}{24 hours} \\
    \cline{2-7}
    & Scope of simulation & \multicolumn{5}{c}{Singapore network with 6.5M agents} \\
\hline
    \multirow{4}{*}{\makecell{Scenario \\ factor}} & Modal availability
    & Existing Modes & \multicolumn{4}{c}{Existing Modes + AMOD} \\
    \cline{2-7}
    & Num. of trips\tnote{a} & 9,656,986 & 9,661,473 & 9,658,097 & \multicolumn{2}{c}{9,660,855} \\
    \cline{2-7}
    & Fleet size\tnote{b} & - & 43,000 & 33,000 & \multicolumn{2}{c}{27,000} \\
    \cline{2-7}
    & Fleet composition & - & \multicolumn{4}{c}{4- and 6-seaters} \\
\hline
\end{tabular}
\begin{tablenotes}
    \small
    \item[a] This total number of trips include 665,929 freight trips across all scenarios.
    \item[b] Fleet size taken from \cite{oh2020amod_tra}.
\end{tablenotes}
\end{threeparttable}
\end{table*}

\section{Results and Analysis}\label{sec:SimResult}
\subsection{MFD: Analysis, Modeling, and Estimation}\label{sec:Result_MFD}
The \textit{Supply} module simulates multimodal network performance (travel demand from the pre-day and within-day models) and specifically, all modes listed in Table \ref{Table_TripSupportModes}. For our analysis, the modes have been classified into two categories, based on whether they contribute to vehicle (\textit{vMFD}) and passenger flow (\textit{pMFD}) respectively. First, the private vehicle trips ($PVT$) contribute to both passenger and vehicle traffic on the network. In the case of on-demand services, $MOD$ and $AMOD$ contribute to both categories when the service vehicle drives with passenger(s). In contrast, $MOD_{OP}$ and $AMOD_{OP}$ represent  operational movements, including empty trips to pick up the passenger, cruising for parking or moving to a parking location, and hence, contribute to only vehicle traffic. Public transit passenger trips are captured by the modes $Bus$ (or $Rail$) at the passenger level, while $Bus_{OP}$ represents the bus vehicle movement with fixed routes and schedules. Also note that all trains (labeled as $Rail_{OP}$) are operated on the rail network and do not directly affect road network traffic. Other modes (labeled as $Other$) were also considered, such as $walking$, for passenger flow estimation. As noted in Section \ref{sec:SecScenario}, the freight commodity flow is considered through background freight traffic and accounted for in the vehicle flow estimation.

\begin{table*}[ht]
\centering
\begin{threeparttable}
\caption{Travel Modes}\label{Table_TripSupportModes}
\begin{tabular}{l l | c | c}
\hline
\textbf{\makecell{Category}} & \textbf{\makecell{Mode}} & \textbf{\makecell{Vehicle flow \\ (\textit{vMFD})}} & \textbf{\makecell{Passenger flow \\ (\textit{pMFD})}} \\
\Xhline{1.5pt}
\multirow{2}{*}{\makecell{PVT}} & $Car/Carpool$ & \checkmark & \checkmark \\ 
                               & $Taxi$ & \checkmark & \checkmark \\
\hline
\multirow{2}{*}{MOD} & $MOD$ & \checkmark & \checkmark \\ 
                     & $MOD_{OP}$\tnote{a} & \checkmark & - \\
\hline
\multirow{2}{*}{AMOD} & $AMOD$ & \checkmark & \checkmark \\ 
                      & $AMOD_{OP}$\tnote{a} & \checkmark & - \\
\hline
\multirow{4}{*}{\makecell{PT}} & $Bus$ & - & \checkmark \\ 
                                     & $Bus_{OP}$\tnote{b} & \checkmark & - \\
                                     & $Rail$ & - & \checkmark \\
                                     & $Rail_{OP}$\tnote{c} & - & - \\
\hline
\multirow{1}{*}{Other} & & - & \checkmark \\
\hline
\multirow{1}{*}{Freight} & & \checkmark & - \\ 
\hline
\end{tabular}
\begin{tablenotes}
    \footnotesize
    \item[a] $MOD_{OP}$/$AMOD_{OP}$ represents empty trips made by MOD and AMOD service vehicles for  operational purposes (such as driving to passenger, parking, cruising).
    \item[b] Travel details on $Bus_{OP}$ is collected from the bus trajectory with the pre-defined lines and frequency.
    \item[c] Trains are operated in an underground rail network ($Rail_{OP}$) and excluded from both levels.
\end{tablenotes}
\end{threeparttable}
\end{table*}

\begin{figure}[ht]
 \begin{minipage}{0.5\textwidth}
  \centering
    \includegraphics[scale=0.45]{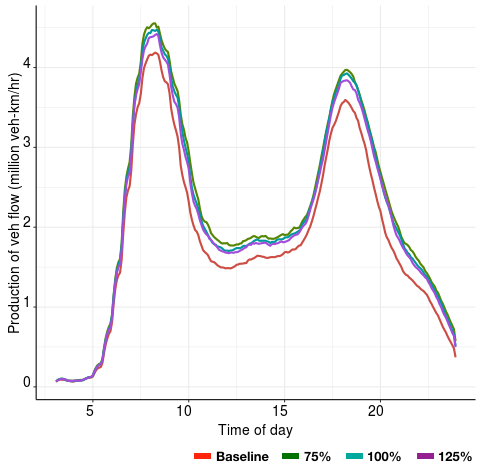}
	\subcaption{Simulated $P_V$}\label{fig:Fig_DistVehFlow_Pv}
\end{minipage}
~\hfill
 \begin{minipage}{0.5\textwidth}
  \centering
    \includegraphics[scale=0.45]{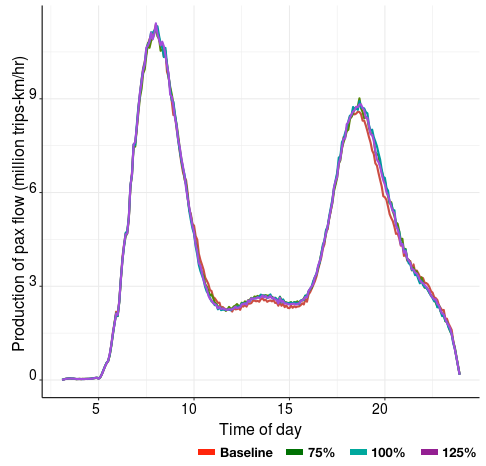}
	\subcaption{Simulated $P_P$}\label{fig:Fig_DistVehFlow_Pp}
  \end{minipage}
\caption{Distribution of Network Production over Time-of-day}\label{fig:Fig_DistProduction}
\end{figure}

Figure \ref{fig:Fig_DistProduction} presents the temporal distribution of network-wide production of vehicle ($\mathcal{P}_V$) and passenger flow ($\mathcal{P}_P$). At the vehicle level (Figure \ref{fig:Fig_DistVehFlow_Pv}), one can notice that traffic flow increases significantly from the baseline scenario with the introduction of AMOD, especially during the peak periods. Moreover, in the lower pricing scenarios, which require a larger fleet size to accommodate the higher AMOD demand (Table \ref{Table_Settings}), we observe increased traffic flows than in the higher pricing case (125\% scenario). 
In contrast, unlike vehicle production, passenger production curves (Figure \ref{fig:Fig_DistVehFlow_Pp}) do not change significantly across scenarios, indicating that the temporal distribution of passenger flows is not significantly affected by the increased traffic flows on the network. 

\begin{figure}[ht]
 \begin{minipage}{0.5\textwidth}
  \centering
    \includegraphics[scale=0.46]{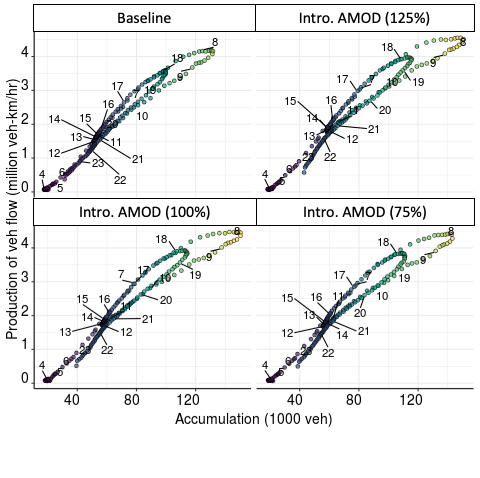}
	\subcaption{Simulated $\mathcal{P_V}$}\label{fig:Fig_vMFD_Est}
\end{minipage}
~\hfill
 \begin{minipage}{0.5\textwidth}
  \centering
    \includegraphics[scale=0.46]{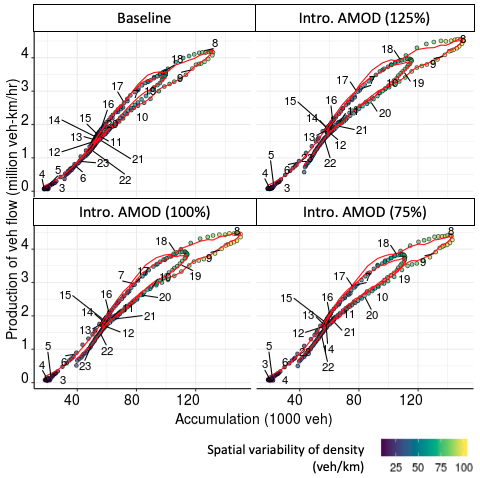}
	\subcaption{Predicted $\mathcal{P_V}$ (Red)}\label{fig:Fig_vMFD_Pre}
  \end{minipage}
\caption{\textit{vMFD}: $\mathcal{P}_V = f(\mathcal{A}_V, \gamma)$}\label{fig:Fig_vMFD}
\end{figure}

Figure \ref{fig:Fig_vMFD_Est} plots the \textit{vMFD}, which relates the production of vehicle traffic ($\mathcal{P}_V$) with vehicle accumulation ($\mathcal{A}_V$) and spatial variability of density ($\gamma$). The time-of-day is also marked on each point of production/accumulation in the figure. Two distinct patterns are visually identifiable, showing the loading and unloading of traffic congestion before and after AM and PM peak periods. Comparing the scenarios, the maximum accumulation of vehicles during the peak increases by 8.7--14.5\% in the AMOD scenarios (150,778, 150,274, 143,155 vehicles for the 75\%, 100\%, and 125\% scenario respectively) from that of baseline (131,689 vehicles). In the case of vehicle production, maximum production increases by about 5.6--8.8\% from the baseline to AMOD scenarios: 4,186,462 veh-km/hr (Baseline), 4,553,106 veh-km/hr (75\% pricing), 4,474,012 veh-km/hr (100\% pricing), and 4,419,385 veh-km/hr (125\% pricing). 
The heterogeneity of network congestion also increases in the AMOD scenarios: the maximum spatial variability of density ($\gamma$) increases from 88 (veh/km) in the baseline to 97--102 (veh/km) in the AMOD scenarios at around 8AM (morning peak period). This increase in heterogeneity leads to the appearance of clockwise hysteresis loops in the \textit{vMFD}, which demonstrate the delay in the recovery of production from the congested state. We quantify the magnitude of hysteresis (\cite{geroliminis2011hysteresis}) by the gap between the production values when loading ($\mathcal{P}_V^l$) and unloading ($\mathcal{P}_V^u$) at a given accumulation level as:

\begin{equation}\label{eq:eq_hys}
h(\mathcal{A}_V) = \Delta \mathcal{P}(\mathcal{A}_V) = \mathcal{P}_v^{l} (\mathcal{A}_V) - \mathcal{P}_v^{u} (\mathcal{A}_V)
\end{equation}

\begin{figure}[ht]
\centering
 \includegraphics[scale=0.5]{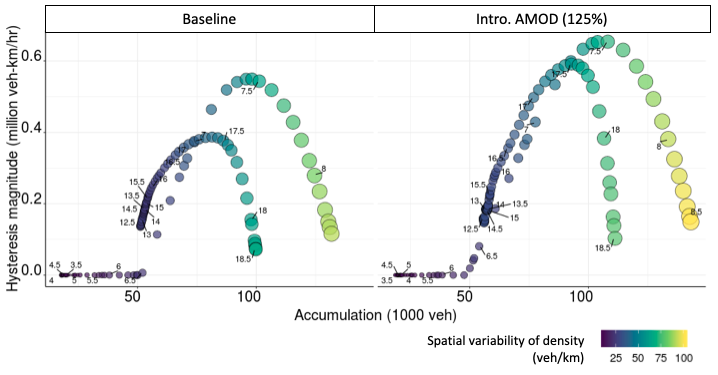}
\caption{Magnitude of Hysteresis ($h(\mathcal{A}_V)$)}\label{fig:Fig_vMFD_Hys}
\end{figure}

Note that in computing the hysteresis, we have used a smoothing spline estimate (\cite{kimeldorf1970correspondence}) to interpolate the production values where required.
Figure \ref{fig:Fig_vMFD_Hys} compares the magnitude of hysteresis between the baseline and the 125\% pricing scenario. In the baseline, the maximum value is 549,065 and 386,942 (veh-km/hr) during the AM and PM peak period respectively. In the AMOD scenario, $h(A_V)$ increases to 649,216--653,581 and 547,930--591,355 (veh-km/hr) for the two peak periods. The total hysteresis during AM and PM peak period ($\mathcal{H} = \int_{t=1}^{T}h(\mathcal{A}_V)dt$) increases by around 24.49--28.56\% when introducing the AMOD service.

According to Eq.\ref{eq:eq_vmfd_hete}, the shape of the MFD is determined by the two variables ($\mathcal{A}_V$, $\gamma$) and model parameters ($a, b, c, d, r$). We estimate the parameters using a nonlinear least squares method (\cite{kass1990nonlinear}) to fit the simulated data ($\mathcal{P}_V^{'}$) with constraints on production $\mathcal{P}_v $($\geqslant0$), accumulation $\mathcal{A}_V$ ($0 \leqslant \mathcal{A}_v \leqslant max(\mathcal{A}_V^{'})$) and space-mean speed $\mathcal{S}(\forall v\in\mathcal{V}: \partial \mathcal{S_{V}}/ \partial \mathcal{A}_v\leq 0)$, where $v\in\mathcal{V}$ (set of road-based modes).

\begin{equation}\label{eq:eq_fitting_mfd}
\min_{a, b, c, d, r}\mathbf{Z}= \| \mathcal{P}_V-\mathcal{P}_{V}^{'} \| ^2
\end{equation}

Table \ref{Table_MFD_Est} lists the estimated parameters, which were all found to be statistically significant. The predicted vehicle production curve (based on the fitted model) for each scenario is shown by the red line in Figure \ref{fig:Fig_vMFD_Pre}, which illustrates the evolution of network dynamics by time-of-day and captures the hysteresis loops during the on- and off-set of congestion. The discrepancy between the simulated and predicted production is measured using the normalized root mean square error (RMSN) in Eq. \ref{eq:eq_rmsn}, and ranges between 0.034--0.036\% over the scenarios. 
 
\begin{equation}\label{eq:eq_rmsn}
RMSN=\frac{\sqrt{T \sum_{t=1}^{T}[\mathcal{P}_V(t) - \mathcal{P}_V^{'}(t)]^2}}{\sum_{t=1}^{T}\mathcal{P}_V^{'}(t)}
\end{equation}

\begin{figure}[ht]
 \begin{minipage}{0.5\textwidth}
  \centering
    \includegraphics[scale=0.46]{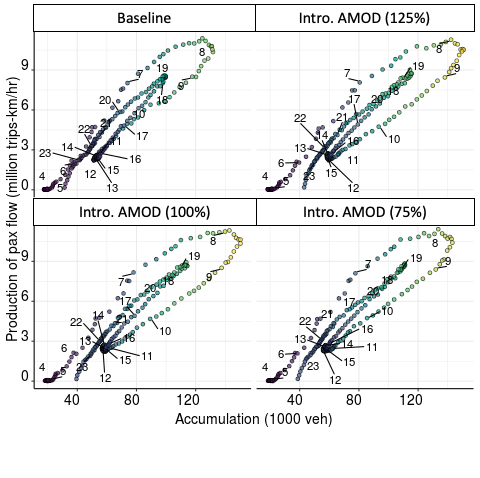}
	\subcaption{Simulated $\mathcal{P_P}$}\label{fig:Fig_pMFD_Est}
\end{minipage}
~\hfill
 \begin{minipage}{0.5\textwidth}
  \centering
    \includegraphics[scale=0.46]{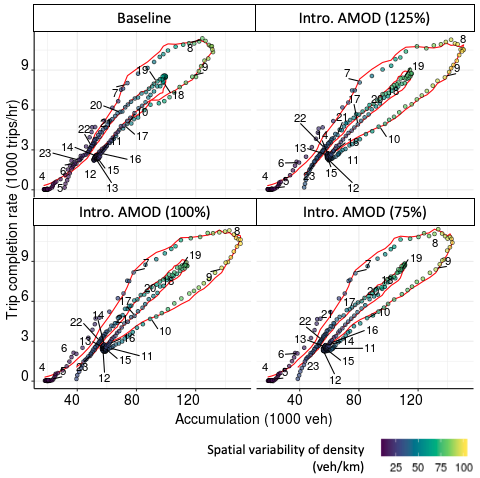}
	\subcaption{Predicted $\mathcal{P_P}$ (Red)}\label{fig:Fig_pMFD_Pre}
  \end{minipage}
\caption{\textit{pMFD}: $\mathcal{P}_P = f(\mathcal{A}_V, \gamma)$}\label{fig:Fig_pMFD}
\end{figure}

In case of the \textit{pMFD}, Figure \ref{fig:Fig_pMFD_Est} shows the production of passenger flow with respect to the aggregate number of vehicles on the network and the spatial variability of density. The shape of the \textit{pMFD} is different from that observed in the case of the \textit{vMFD}. It shows (i) a larger gap between two production curves of loading and unloading during the AM peak (resulting in large clockwise hysteresis loops), and (ii) small counter-clockwise hysteresis loop during the PM peak. These two points can be attributed to the nature of passenger trip distances as elaborated below:

\begin{enumerate}[label=(\roman*)]
    \item Difference in the average trip distances at the vehicle and passenger level ($TD_V > TD_P$). The average trip distance of vehicle ($TD_V$) reduces from around 12.5km (while loading) to 10--11km (while unloading after 8:30AM). In case of $TD_P$, it decreases more significantly from around 9km (while loading) to 6.5km (while unloading). Since the production is determined by both trip completion rate and trip distance, the larger decrease in $TD_P$ results in a higher trip completion rate, as well as a larger gap of $\mathcal{P}_p$ between the loading and unloading in case of the \textit{pMFD}.
    \item Longer trip distances while unloading during the PM peak period. The passenger trip distance ($TD_P$) appears to be longer than 8km after 7PM, during the unloading, while being shorter (7--8km) for those trips completed before 7PM, during the loading. This contributes to higher production during unloading and results a counter-clockwise hysteresis loop. Additional clues can be found in the temporal demand pattern by activity types (see Section \ref{sec:SecScenario}): more trips (e.g. \textit{Other} activity in Figure \ref{fig:Fig_Demand_Type}) are generated and contribute to higher production in the offset of congestion during the PM peak period.
\end{enumerate}

In a similar manner as the $vMFD$, we estimate the model described in Eq.\ref{eq:eq_pmfd_hete} and the estimated parameters are summarized in Table \ref{Table_MFD_Est}, all of which were found to be statistically significant. The discrepancy between simulated and predicted passenger productions (quantified by the RMSN) are found to range between 0.074--0.079 \% across the scenarios.
Also, as shown in Figure \ref{fig:Fig_DistVehFlow_Pp} and Figure \ref{fig:Fig_pMFD_Est}, the maximum and overall temporal patterns of passenger production ($\mathcal{P}_P$) remain similar across the scenarios, in contrast with the the distinct impacts on $\mathcal{P}_V$ in the $vMFD$ observed with the introduction of AMOD. This may be ascribed to a range of factors, one of which is the cannibalization of transit by AMOD (explained in Section \ref{sec:SecScenario}). Even though the road network congestion is more severe in the AMOD scenarios (as verified in Section \ref{sec:Result_Effects_Congestion}), the effects of network congestion on the production of passenger flow may be minimal as a significant share of AMOD (`faster' modes in general but which are affected by the additional network congestion) includes shifts from transit (`slower' modes in general but which are unaffected by network congestion). 

\begin{adjustbox}{width=1\textwidth,center=\textwidth}
\centering
\begin{threeparttable}
\caption{Estimation Result for \textit{MFD}}\label{Table_MFD_Est}
\begin{tabular}{c l r r r r r r}
\hline
    \multirow{2}{*}{\textbf{Model}} & \multirow{2}{*}{\textbf{}} & \multicolumn{6}{c}{\textbf{Parameters}} \\
    \cmidrule(r){3-8}
    & & $a$ & $b$ & $c$ & $d$ & $\rho$ & $r$ \\
\Xhline{1.5pt}
    \multirow{4}{*}{\textit{vMFD}} & Baseline & $0.284$ & $7.50 \cdot 10^{-5}$ & $-6.28 \cdot 10^{-10}$
 & $1.954 \cdot 10^{-15}$ & - & $-0.01346$\\
    \cline{2-8}
     & 75\% & $0.328$ & $6.65 \cdot 10^{-5}$ & $-4.79 \cdot 10^{-10}$
 & $1.286 \cdot 10^{-15}$ & - & $-0.01462$ \\
    \cline{2-8}
     & 100\% & $0.366$ & $6.26 \cdot 10^{-5}$ & $-4.38 \cdot 10^{-10}$
 & $1.151 \cdot 10^{-15}$ & - & $-0.01465$ \\
    \cline{2-8}
     & 125\% & $0.298$ & $7.10 \cdot 10^{-5}$ & $-5.42 \cdot 10^{-10}$
 & $1.537 \cdot 10^{-15}$ & - & $-0.01452$ \\
\hline
\multirow{4}{*}{\textit{pMFD}} & Baseline & $0.608$ & $6.42 \cdot 10^{-5}$ & $-8.99 \cdot 10^{-10}$ & $2.94 \cdot 10^{-15}$ & $5.22 \cdot 10^{-6}$ & $0.00634$ \\
    \cline{2-8}
     & 75\% & $0.734$ & $5.07 \cdot 10^{-5}$ & $-5.73 \cdot 10^{-10}$ & $1.654 \cdot 10^{-15}$ & $5.39 \cdot 10^{-6}$ & $-0.00705$ \\
    \cline{2-8}
     & 100\% & $0.662$ & $5.48 \cdot 10^{-5}$ & $-6.12 \cdot 10^{-10}$ & $1.794 \cdot 10^{-15}$ & $4.84 \cdot 10^{-6}$ & $-0.00534$ \\
    \cline{2-8}
     & 125\% & $0.622$ & $5.70 \cdot 10^{-5}$ & $-6.85 \cdot 10^{-10}$ & $2.04 \cdot 10^{-15}$ & $5.19 \cdot 10^{-6}$ & $-0.00207$ \\
\hline
\end{tabular}
\end{threeparttable}
\end{adjustbox}

\vspace{1\baselineskip}
\begin{adjustbox}{width=1\textwidth,center=\textwidth}
    \centering
    \begin{threeparttable}
    \caption{Primary (Well-to-wheels) Energy Consumption (unit: kWh)}\label{Table_Env_Energy}
    \begin{tabular}{l || c  c  c  c  c | c || c  c | c }
    \hline
        \multirow{3}{*}{\textbf{Scenarios}} & \multicolumn{6}{c||}{Fuel} & \multicolumn{3}{c}{Electricity} \\
        & \multicolumn{1}{c}{$v=PVT$} & \multicolumn{1}{c}{$Bus_{OP}$} & \multicolumn{1}{c}{$MOD$} & \multicolumn{1}{c}{$MOD_{OP}$} & \multicolumn{1}{c|}{$Freight$} & \multicolumn{1}{c||}{Total} & $AMOD$ & $AMOD_{OP}$ & Total\\
    \Xhline{1.5pt}
        Baseline & 10,186,083 & 505,332 & 2,901,130 & 1,337,413 & 3,798,208 & 18,728,167 & 0 & 0 & 0 \\
        75\% & 9,398,493 & 503,049 & 2,254,457 & 1,003,054 & 3,665,271 & 16,824,324 & 4,107,287 & 2,354,117 & 6,461,405 \\
        100\% & 9,596,024 & 503,917 & 2,394,551 & 1,098,965 & 3,663,389 & 17,256,844 & 3,344,555 & 1,905,370 & 5,249,925 \\
        125\% & 9,693,896 & 504,227 & 2,477,469 & 1,154,828 & 3,661,884 & 17,492,304 & 2,815,517 & 1,593,608 & 4,409,125 \\
    \hline
    \end{tabular}
    \end{threeparttable}
\end{adjustbox}

\vspace{1\baselineskip}
\begin{adjustbox}{width=1\textwidth,center=\textwidth}
    \centering
    \begin{threeparttable}
    \caption{Vehicle Emission: $NO_{x}$ and $PM$ (unit: kg)}\label{Table_Env_Emission}
    \begin{tabular}{l || c c | c c | c c | c c | c c | c c }
    \hline
        \multirow{2}{*}{\textbf{Scenarios}}& \multicolumn{2}{c}{$v=PVT$} & \multicolumn{2}{c}{$Bus_{OP}$} & \multicolumn{2}{c}{$MOD$} & \multicolumn{2}{c}{$MOD_{OP}$} & \multicolumn{2}{c}{$Freight$} & \multicolumn{2}{c}{Total} \\
        & $NO_{x}$ & $PM$ & $NO_{x}$ & $PM$ & $NO_{x}$ & $PM$ & $NO_{x}$ & $PM$ & $NO_{x}$ & $PM$ & $NO_{x}$ & $PM$ \\
    \Xhline{1.5pt}
        Baseline & 1080.6 & 72.3 & 963.4 & 17.9 & 272.9 & 20.6 & 125.9 & 9.5 & 2856.8 & 63.1 & 5299.7 & 183.4 \\
        75\% & 993.1 & 66.7 & 954.1 & 17.8 & 209.8 & 16.0 & 93.3 & 7.1 & 2745.9 & 60.6 & 4996.1 & 168.3 \\
        100\% & 1015.4 & 68.1 & 956.2 & 17.9 & 223.5 & 17.0 & 102.6 & 7.8 & 2746.5 & 60.7 & 5044.2 & 171.4 \\
        125\% & 1025.7 & 68.8 & 958.2 & 17.9 & 231.6 & 17.6 & 108.1 & 8.2 & 2744.5 & 60.6 & 5068.1 & 173.1 \\
    \hline
    \end{tabular}
    \end{threeparttable}
\end{adjustbox}


\subsection{Impacts on Energy, Emissions and Congestion}\label{sec:Result_Effects}
\subsubsection{Energy and Emissions}\label{sec:Result_Effects_Environment}
In this section, we examine the impacts of AMOD on energy and emissions at the network level. We assume that the AMOD fleet is fully composed of battery electric vehicles (BEV) and the other vehicle categories are composed of gasoline/diesel-fueled vehicles (Euro 6 standard for passenger vehicles, bus, and freight trucks). Table \ref{Table_Env_Energy} and Table \ref{Table_Env_Emission} summarize the emissions and energy consumption for each travel mode ($v$) based on the total vehicle-km traveled (VKT). Note that this VKT is equivalent to the total $\mathcal{P}_V$ for 24h, which is 31.78, 37.65, 36,65, and 35.51 million-km for the baseline, 75\%, 100\%, and 125\% scenarios respectively.
As noted previously, we observe a significant increase in VKT ranging from 11.8-18.5\% for the AMOD scenarios, compared to the baseline.
 
Energy consumption of the AMOD fleets is measured using an average energy consumption rate (ECR). According to real-world estimation data \citep{fetene2014report}, the ECR decreases with vehicle travel distance as follows: 233Wh/km, 183Wh/km, 166Wh/km for short ($TD_{v} \leq 2km$), medium ($2km \leq TD_{v} \leq10km$), and long distances ($TD_{v} \geq 10km$). The energy consumption is computed by multiplying the production factor (2.99, US average energy-to-fuel ratio), which incorporates well-to-wheels effects while taking into account the transmission and distribution losses of BEVs. Accordingly, the total energy consumption is 6.46GWh, 5,25GWh, 4,41GWh for the 75\%, 100\%, 125\% scenarios respectively.
As anticipated, the increase in VKT, in lower pricing scenarios, results in larger energy consumption for both service and operational purposes. Note that a significant portion of energy consumption is caused by the operating trips (empty trips for passenger pick-up, cruising, parking) taking around 36\% of total energy consumption across AMOD scenarios. 
Further, for the existing road-based modes (non-electric vehicles), we compute energy consumption using the miles per gallon gasoline equivalent (MPGe) of each vehicle type. By assuming the future MPGe as 47(5.0L/100km) and 52(4.5L/100km) for gasoline and diesel-powered vehicles respectively (\cite{EU2011Emission}), total consumption (by $PVT$, $Bus_{OP}$, $MOD$, $MOD_{OP}$, $Freight$) is determined to be 18.73GWh, 16.82GWh, 17.26GWh, and 17.49GWh for the four scenarios respectively.
Note that 1 MPGe is equivalent to 0.04775km/kWh (\cite{EPA2011Emission}) and the corresponding average energy-to-fuel ratios are 1.17 and 1.05 for gasoline and diesel respectively. Thus, total energy consumption by all vehicles increases with the introduction of AMOD by 24.33\%, 20.18\%, and 16.94\% for the 75\%, 100\%, and 125\% scenarios respectively.

In the case of vehicle emissions, we consider the production of $NO_x$ and \textit{PM} (exhaust particulate matter) by passenger cars as well as buses and trucks on the network. According to the emission testing (Euro-6 standard) results in \cite{oecdreport2017_emission}, the unit emissions for $NO_x$ and $PM$ are estimated dependent on vehicle types (passenger cars, buses, trucks) and congestion as: 0.043--0.063g/km ($NO_x$), 0.0037g/km ($PM$) for passenger car (petrol), 0.69--1.11g/km ($NO_x$), 0.015g/km ($PM$) for buses, 0.28--0.44g/km ($NO_x$), 0.0061--0.010g/km ($PM$) for trucks. Total emissions reduce with the introduction of AMOD from 5,299.7kg ($NO_x$) and 183.4kg ($PM$) in baseline to 4,996--5,068kg ($NO_x$) and 168--173kg ($PM$) in the AMOD scenarios. In summary, the introduction of AMOD may bring about significant emission reductions (4.3--5.7\% in $NO_x$ and 5.6--8.2\% in $PM$), while resulting in more energy consumption (up to 24.33\% from the baseline scenario).

\subsubsection{Congestion and Delay}\label{sec:Result_Effects_Congestion}
\begin{figure}[ht]
 \begin{minipage}{0.5\textwidth}
  \centering
    \includegraphics[scale=0.46]{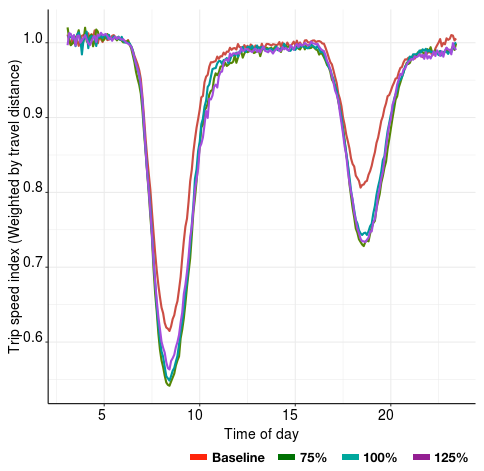}
	\subcaption{Trip Speed Index (TSI)}\label{fig:Fig_TSI}
\end{minipage}
~\hfill
 \begin{minipage}{0.5\textwidth}
  \centering
    \includegraphics[scale=0.46]{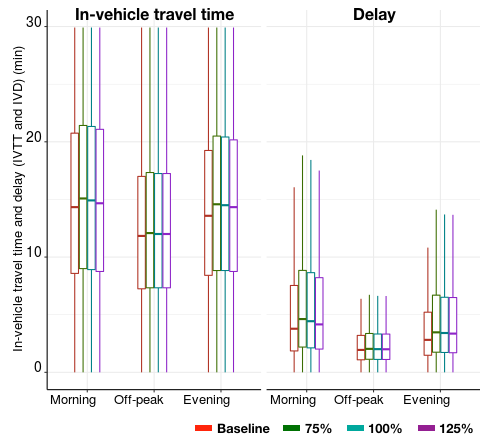}
	\subcaption{$IVTT$ and $IVD$}\label{fig:Fig_IVTT}
  \end{minipage}
\caption{Congestion Effects}\label{fig:Fig_TravelExperience}
\end{figure}

The increase in network traffic contributes to congestion and travel delays. In order to further quantify network congestion, we examine the distance weighted trip speed index (TSI) using the individual vehicle's trip speed ($TS_{\iota}$) and the travel distance from origin to destination ($TD_{\iota}$):

\begin{equation}\label{eq:eq_tsi_weighted}
TSI = \frac{\sum_{\iota}\left(TD_{\iota}*(TS_{\iota}/TS_{\iota}^{0})\right)}{\sum_{\iota}TD_{\iota}}
\end{equation}

where, $TS_{\iota}^{0}$ is the free-flow speed between origin and destination of individual $\iota$. Clearly, as seen in Figure \ref{fig:Fig_TSI}, the trip speed index decreases from 1 in the off-peak period (free-flow) to values of around 0.65 and 0.8 in the AM and PM peak periods respectively. Furthermore, the TSI for AMOD scenarios decreases significantly during the peak periods by 8--11.9\% (AM) and 7.8--9.7\% (PM) from the baseline.  

The increase in VKT and decrease in network speed, as expected, affect travel experience. We quantify this effect using a measure of delay in travel-time ($IVD_{\iota}$) at the individual level ($\iota$):

\begin{equation}\label{eq:eq_delay_ivtt}
IVD_{\iota}=IVTT_{\iota}-IVTT_{\iota}^{0}
\end{equation}

where, $IVTT_{\iota}$ is the individual in-vehicle travel-time; $IVTT_{\iota}^{0}$ is the free-flow travel time 
The distributions of $IVTT$ and $IVD$ are shown in Figure \ref{fig:Fig_IVTT}. Compared to the baseline (5.2 and 3.7min of $IVD$ for AM and PM peak period), $IVD$ increase ranges from 7.8--15\%, 20--23\% for AM and PM peak periods across the AMOD scenarios. 
Waiting times are around 4.4 min (off-peak) and range from 5--6 min (during peak periods) on average (ranging from 1 to 3 min of delay).

\section{Conclusion}\label{sec:Conclusion}
This paper evaluates the network impacts of AMOD from an MFD perspective utilizing agent-based simulation. The simulation framework models activity-based travel demand, supply (including fleet operations and multimodal network performance) and their interactions.  
Scenario simulations of the entire urban network of Singapore yield several insights into the impacts of AMOD:
Introduction of AMOD services may induce additional vehicle traffic resulting in more congestion relative to the baseline scenario. The network congestion in AMOD scenarios is due in part to the demand patterns (i.e. cannibalization of transit shares) as well as dead-heading and empty trips for operational purposes. The vehicle accumulation and production increases by 8.7--14.5\% and 5.6--8.8\% respectively, and the total magnitude of hysteresis loops increases by more than 24\% with the introduction of the AMOD service. Despite the increase in network congestion, the passenger production is not significantly impacted. The estimated models of \textit{vMFD} and \textit{pMFD} predict the production at the vehicle and passenger level and their dynamics accurately. 
In addition, the impacts of AMOD in terms of energy and emissions is quantified. The introduction of AMOD leads to increased energy consumption (by 16.94--24.33\% from baseline), although vehicle emissions in terms of $NO_x$ and $PM$ are reduced (by 4.3--5.7\% and 5.6--8.2\%, respectively). Moreover, the travel delay has been increased up to 23\% in the case of the AMOD scenario.

Based on the simulation and modeling framework, several avenues for future research remain, including the testing of (existing/emerging) MFD-based and other traffic management measures and policies (i.e. vehicle quota systems, route guidance systems, perimeter control, congestion pricing in multimodal urban networks) for maximizing social welfare at both the local and urban scale. The proposed framework can also be applied to evaluate the effect of long-term impacts of AMOD on land-use as well as car-ownership, which are interesting areas for future research.

\section*{Acknowledgements}
This research is supported in part by the Singapore Ministry of National Development and the National Research Foundation, Prime Minister’s Office under the Land and Liveability National Innovation Challenge (L2 NIC) Research Programme (L2 NIC Award No L2NICTDF1-2016-4). Any opinions, findings, and conclusions or recommendations expressed in this material are those of the author(s) and do not reflect the views of the Singapore Ministry of National Development and National Research Foundation, Prime Minister’s Office, Singapore. Also, this research was supported by Basic Science Research Program through the National Research Foundation of Korea (NRF) funded by the Ministry of Education (NRF-2019R1A6A3A12031439).


\bibliography{References}

\end{document}